\documentstyle[preprint,epsfig,aps,pre]{revtex}

\begin{document}
\title{Field dependent collision frequency of the two-dimensional driven random Lorentz gas}
\author{Christoph Dellago\footnote{Corresponding Author. {\em Email address}: dellago@chem.rochester.edu}}
\address{Department of Chemistry, University of Rochester, Rochester, NY 14627, USA}
\author{Henk v. Beijeren}
\address{Institute for Theoretical Physics, University of Utrecht,
Postbus 80006, Utrecht 3508 TA, The Netherlands} 
\author{Debabrata Panja}
\address{Instituut Lorentz, Universiteit Leiden, Postbus 9506, 2300 RA Leiden, The Netherlands}
\author{J. R. Dorfman}
\address{Institute for Physical Sciences and Technology and Department of Physics, University of Maryland, 
College Park, MD 20742, USA}
\date{\today}
\maketitle


\begin{abstract}
In the field-driven, thermostatted Lorentz gas the collision frequency 
increases with the magnitude of the applied field due to long-time
correlations.  We study this effect with computer simulations and confirm the 
presence of non-analytic terms in the field dependence of the collision
frequency as predicted by kinetic theory.
\end{abstract}


\newpage

In a recent paper \cite{PDvB}, Panja, Dorfman and van Beijeren used kinetic
theory to derive analytic expressions for the Lyapunov exponents of the 
random, two-dimensional field-driven Lorentz gas at moderately high densities. Their 
approach is based on the BBGKY hierarchy equations and takes into 
account correlated collision sequences. Such "ring collisions" lead to 
long-time tails in the Green-Kubo time correlations functions affecting
the transport properties of the system \cite{kintheory}. Panja {\em et al.} 
studied the effect of ring collisions on the Lyapunov exponents and found that 
long-time correlations cause a logarithmic dependence of the Lyapunov 
exponents on the applied field strength. These non-analytic terms can 
be traced back to logarithmic terms in the field dependence of the 
collision frequency. Unfortunately, the non-analytic contributions to the 
Lyapunov exponents are too small to be detected with current computer 
simulation techniques \cite{DP97}. In contrast, the predicted effect on the collision
frequency is large enough to be observed numerically. The purpose of this 
note is to verify the presence of logarithmic terms in the collision 
frequency with computer simulations.

The random, two-dimensional field-driven Lorentz gas, shown in Fig. 
\ref{fig:geometry}, consists of a point particle of mass $m$ and 
charge $q$ moving under the influence of an external homogeneous 
field ${\bf E}$ in a two-dimensional array of circular non-overlapping
scatterers with radius $a$. The scatterers are fixed at random 
positions in the plane and have a density of $n=N/A$, 
where $N$ is the total number of scatterers in the area $A$. 
In our numerical simulations we use periodic boundary conditions as 
indicated in Fig. \ref{fig:geometry}. When the point 
particle leaves the simulation box through one specific boundary, it 
re-enters it through the opposite boundary. Between collisions with 
the scatterers the point particle moves smoothly according to
\begin{equation}
\label{equ:motion}
\dot {\bf r}=\frac{\bf p}{m}, \qquad \dot{\bf p} = q{\bf E}-\alpha {\bf p}, 
\qquad \alpha=q\frac{{\bf E} \cdot {\bf p}}{p^2},
\end{equation}
where ${\bf r}\equiv\{r_x, r_y\}$ and ${\bf p}\equiv\{p_x, p_y \}$ are the 
position and the momentum of the moving particle, respectively. The second 
term in the momentum space part of the equations of motion is a Gaussian thermostat 
designed to remove the dissipated energy and keep the speed $v=p/m$ of the 
moving particle constant\cite{Moran}. When the point particle collides with a scatterer 
it is reflected elastically, i.e.,
\begin{equation}
{\bf v}_+={\bf v}_- - 2({\bf v}_-\cdot {\bf u}),
\end{equation}
where ${\bf u}$ is the unit vector in the direction from the center of the 
scatterer to the collision point and ${\bf v}_-$ and ${\bf v}_+$ are the 
pre- and post-collisional velocities of the moving particles respectively. 

In equilibrium, i.e. for $E=0$, and for long times, the phase distribution is
uniform on the energy shell. In this case the collision frequency of the point 
particle with the scatterers can be obtained by determining the fraction
of phase space available for collision in an infinitesimal time 
interval \cite{Zwanzig}:
\begin{equation}
\nu=\frac{2nav}{1-\pi n a^2}.
\end{equation}
In the presence of an external perturbation, 
however, the phase space distribution of non-equilibrium steady state 
collapses onto a multifractal strange attractor with information 
dimension strictly less than the phase space dimension. As a consequence,
no analytical expression for the phase space distribution is available
and the calculation of phase space integrals becomes cumbersome. 
Nevertheless, Panja, Dorfman and van Beijeren succeeded 
in determining the effect of the external field on the collision 
frequency by adopting a kinetic theory approach \cite{PDvB}.

Qualitatively, it is evident that the external field increases the 
collision frequency. In the presence of the field, the velocity of the particle
tends to be aligned in field direction before a collision. Immediately
after the collision the velocity of the particle will therefore 
point against the field. Due to the action of the field the particle
will turn around and possibly hit the same scatterer again 
after a time shorter than the average equilibrium collision time.
The typical time scale for such a reorientation is of the order of 
$mv/qE$ \cite{DGP95}. Such correlated collisions enhance the collision frequency.

This argument was made quantitative by Panja, Dorfman, and van Beijeren \cite{PDvB}.
As a consequence of correlated collisions, i.e. multiple collisions of the particle
with the same scatterer, separated by sequences of intermediate collisions, a field 
dependent contribution appears in the collision frequency:
\begin{equation}
\label{equ:rate}
\nu=\frac{2nav}{1-\pi n a^2}+\frac{a\varepsilon^2}{2\pi v}\ln \frac{2nav}{\varepsilon},
\end{equation}
where $\varepsilon = q\left| {\bf E}\right| / (mv)$. The first term on the
right hand side of the above equation is the collision frequency in equilibrium. The 
second, non-analytic field dependent term is responsible for the non-analytic 
dependence of the Lyapunov exponents on the field strength. 

To verify the presence of the logarithmic term in equation (\ref{equ:rate})
we have performed extensive simulations of the driven random Lorentz gas
at various densities and field strengths. The collision frequency is obtained 
in a straightforward way by following the time evolution of the system 
for a long time and counting the number of collisions. For this purpose
we use an analytical solution of equations of motion (\ref{equ:motion})
and determine the collision point and the collision time of the moving 
particle with the scatterers numerically \cite{Moran}. Typically, we study 
systems with $N=10^5$ scatterers in a square simulation box with periodic
boundary conditions. Only at the highest density $(n=0.15a^{-2})$ we use $N=5000$
scatterers. By averaging our results over various scatterer configurations
we integrate over the quenched disorder of the randomly placed scatterers.
Computing time is saved by dividing the simulation box into cells such 
that at each time only a few scatterers need to be considered 
as possible collision partners. For each density and field strength 
we typically carry out a total of more than $10^9$ collisions obtaining 
collision frequencies accurate to better than $0.01\%$. This high accuracy is 
needed to detect the small changes in collision frequency in the weak field regime. 

For analysis it is convenient to rewrite equation (\ref{equ:rate}) as
\begin{equation}
\label{equ:delta_nu}
\frac{\delta \nu}{\varepsilon^2}= \frac{a}{2\pi v} \ln(2na^2) -  
                                \frac{a}{2\pi v}\ln \frac{\varepsilon a}{v} 
\end{equation}
where $\delta \nu$ is the deviation of the collision frequency from its
equilibrium value. Accordingly, $\delta \nu / \varepsilon^2$ should behave
linearly when plotted as a function of $\ln (\varepsilon a / v)$. Figure 
\ref{fig:delta_nu} shows $\delta \nu / \varepsilon^2$ as a function of
$\ln (\varepsilon a / v)$ for the densities $n=0.001a^{-2}$, $0.002a^{-2}$, 
$0.005a^{-2}$, $0.01a^{-2}$, $0.02a^{-2}$, $0.04a^{-2}$, $0.08a^{-2}$,
and $0.15a^{-2}$. In Fig. \ref{fig:delta_nu} densities increase from left to right.
Each data point is obtained as an average over 50 to 500 runs with 
different scatterer configurations. The error bars are estimated from
the variation of the collision frequency in these sets of runs. The solid 
lines connect data points corresponding to the same density and the 
dotted lines are straight lines with slope $-a/(2\pi v)$ fitted to the data
in the low field regime. 

For all densities but the highest one $\delta \nu / \varepsilon^2$
is a linear function of $\ln \varepsilon$ in the low 
field strength range with slope $-a/(2\pi v)$ in agreement with equation 
(\ref{equ:delta_nu}). This confirms the existence of the non-analytic 
field dependent term predicted by kinetic theory.

From Fig. \ref{fig:delta_nu} we can also infer the range of validity 
of the theory in the $n$-$\varepsilon$ plane. Clearly, equation 
(\ref{equ:delta_nu}) breaks down for densities larger than about $n=0.08a^{-2}$.
For densities below $n=0.08a^{-2}$ equation (\ref{equ:delta_nu}) holds 
for field strengths below a certain critical value $\varepsilon_c$. The 
intercept of the fitted lines with the $x$-axis can be used as a 
measure of this critical value. $\varepsilon_c$ is an approximately 
linear function of the density growing from $\varepsilon_c \sim 0.006 v/a$ at 
$n=0.001a^{-2}$ to $\varepsilon_c \sim 0.5 v/a$ at $n=0.08a^{-2}$.

However, the agreement between numerical results and analytical theory is not
perfect even for low densities and weak external fields. While the lines
corresponding to equation (\ref{equ:delta_nu}) have the correct slope, their
intercept with the $y$-axis is lower by about $0.2a/v$ than the value 
obtained numerically. This discrepancy is essentially constant in the 
density range we have studied. It must be due to the approximation of the 
probability of return to a given scatterer by a solution of the diffusion 
equation, as described in the Appendix of Ref.~\cite{PDvB}. In order to check
this in detail one would have to explicitly solve the Lorentz-Boltzmann 
equation in presence of the field and thermostat.

In summary, we have unequivocally detected the appearance of non-analytic
terms in the collision frequency and hence, indirectly, in the Lyapunov 
exponents of the random field driven Lorentz gas. Our numerical results
agree quantitatively with the kinetic theory predictions of Panja, Dorfman, 
and van Beijeren.


\bigskip
\noindent
{\bf Acknowledgements}

J. R. D. thanks the National Science Foundation for support under
Grant No. PHY-98-20824.


\newpage
\begin{figure}
\caption{\label{fig:geometry}
Geometry of the random, field-driven Lorentz gas.}
\end{figure}
\begin{figure}
\caption{\label{fig:delta_nu}
$\delta \nu / \varepsilon^2$ as a function of $\ln (\varepsilon a /v)$ for the 
densities $n=0.001a^{-2}$, $0.002a^{-2}$, $0.005a^{-2}$, $0.01a^{-2}$, 
$0.02a^{-2}$, $0.04a^{-2}$, $0.08a^{-2}$, and $0.15a^{-2}$ (from 
left to right). The dotted lines are straight lines with slope $-a/(2\pi v)$ 
fitted to the data. The linear behavior of $\delta \nu / \varepsilon^2$ in the 
weak field regime indicates the presence of the logarithmic terms
predicted by kinetic theory.}
\end{figure}

\end{document}